\documentclass[12pt,preprint]{emulateapj}
\usepackage{graphicx}

\shorttitle{Detectability of Exoplanets in the Infra-Red}
\shortauthors{Stephen R. Kane \& Dawn M. Gelino}
\slugcomment{Submitted for publication in the Astrophysical Journal}

\begin{document}

\title{Detectability of Exoplanet Periastron Passage in the Infra-Red}
\author{Stephen R. Kane, Dawn M. Gelino}
\affil{NASA Exoplanet Science Institute, Caltech, MS 100-22, 770
  South Wilson Avenue, Pasadena, CA 91125, USA}
\email{skane@ipac.caltech.edu}


\begin{abstract}

Characterization of exoplanets has matured in recent years,
particularly through studies of exoplanetary atmospheres of transiting
planets at infra-red wavelenegths. The primary source for such
observations has been the Spitzer Space Telescope but these studies
are anticipated to continue with the James Webb Space Telescope
(JWST). A relatively unexplored region of exoplanet parameter space is
the thermal detection of long-period eccentric planets during
periastron passage. Here we describe the thermal properties and
albedos of long-period giant planets along with the eccentricities of
those orbits which allow them to remain within the habitable zone. We
further apply these results to the known exoplanets by calculating
temperatures and flux ratios for the IRAC passbands occupied by warm
Spitzer, considering both low and high thermal redistribution
efficiencies from the perspective of an observer. We conclude with
recommendations on which targets are best suited for follow-up
observations.

\end{abstract}



\section{Introduction}
\label{introduction}

As the number of known exoplanets continues to rise at a steady pace,
their diversity appears to only increase. For instance, transiting
exoplanets have allowed us to access properties of exoplanet
atmospheres through observations at Infra-Red (IR) wavelengths during
secondary eclipse and atmospheric absorption during primary transit
\citep{ago10,dem07}. Further opportunities for atmospheric studies
have arisen through the detection of phase variations of such planets
as HD~189733b \citep{knu09a} and HD~149026b \citep{knu09b}. These
phase variations have also been detected for non-transiting planets,
including $\upsilon$~And~b \citep{cro10,har06} and HD~179949b
\citep{cow07}. Significant constraints have been placed on the
planet-to-star flux ratio for HD~217107b through ground-based near-IR
observations by \citet{cub11}.

Studies thus far have largely been directed towards short-period
planets which are expected to have an a priori high effective
temperature. In particular, this exclusivity results from a relatively
high thermal flux required from the planet in order for the signature
to be detectable. The exception to this is the planet HD~80606b, with
a highly eccentric orbit ($e = 0.93$) and a period of $\sim 111$ days,
detected by \citet{nae01}. Subsequent Spitzer observations by
\citet{lau09}, as well as the detection of the secondary eclipse of the
planet, were used to measure the out-of-eclipse variations and
estimate the radiative time constant at 8 microns. As pointed out by
\citet{bar07}, ambiguous measurements of radiative time constants
have prevented a consensus on expected planet-wide flow patterns for
short-period planets.

Recent three-dimensional models of planetary atmospheres
\citep{bar05,dob10,for10,kos07,lew10,mad10,rau10,sho08,sud05,thr10}
have made significant progress in deriving the underlying atmospheric
physics which drive the thermal properties and zonal winds in
exoplanetary atmospheres. Highly eccentric planets such as HD~80606b
provide a means to explore atmospheric properties in a different
regime of orbital parameter space since the heating of the atmosphere
during periastron passage can be sufficient to produce a detectable
signal \citep{cow11a}. The possability of investigating eccentric
planets at optical wavelengths has previously been explored by
\citet{kan10} and \citet{kan11a}. This observing window for periastron
passage is brief and requires a reasonable understanding of the
orbital parameters \citep{kan09}.

This paper describes the predicted thermal changes for exoplanets in
highly eccentric orbits. This study is primarily motivated from an
observers point of view and is mostly concerned with planets which are
not known to produce either a secondary eclipse or primary transit. We
derive analytical expressions for the albedos based upon theoretical
models and calculate the effective temperatures and flux ratios,
taking into account the thermal heat redistribution and radiative time
constant. These calculations are applied to the known exoplanets for
which orbital parameters measured from radial velocity data are
available. We explore the dependencies of the planetary effective
temperatures on eccentricity and orbital period and determine the
percentage of the orbits which are spent in their respective habitable
zones. We finally calculate predicted maximum flux ratios during
periastron passage, discuss the effect of spots on detections, and
propose potential targets for Spitzer and JWST observations.


\section{Planetary Effective Temperatures}

In this section we outline the basic assumptions used to calculate the
planetary effective temperatures throughout the remainder of the paper.
These assumptions have been deliberately introduced to be fairly broad
since the intention is to encompass a variety of planets to produce a
first-order approximation of the global distribution of thermal
signatures and detectability. These were designed in such a way so as
to produce testable limits on the flux ratios at periastron passage.

We begin with the luminosity of the host star, which is approximated
as
\begin{equation}
  L_\star = 4 \pi R_\star^2 \sigma T_\mathrm{eff}^4
\end{equation}
where $R_\star$ is the stellar radius, $T_\mathrm{eff}$ is the stellar
effective temperature, and $\sigma$ is the Stefan-Boltzmann
constant. In cases where the radius of the star is not available from
direct measurements, we estimate the radius from the derived values of
the surface gravity $\log g$ using the relation
\begin{equation}
  \log g = \log \left( \frac{M_\star}{M_\odot} \right) - 2 \log
  \left( \frac{R_\star}{R_\odot} \right) + \log g_\odot
\end{equation}
where $\log g_\odot = 4.4374$ \citep{sma05}.

As described by \citet{knu09b}, we can approximate the effective
temperature of a planet, $T_p$, as a blackbody. This approximation
will deviate slightly from the true temperature depending upon albedo,
atmospheric properties, and internal heating. Assuming that the
atmosphere is 100\% efficient at redistributing heat around the
planet, the planetary equilibrium effective temperature is given by
\begin{equation}
  T_p = \left( \frac{L_\star (1-A)}{16 \pi \sigma r^2}
  \right)^\frac{1}{4}
\end{equation}
where $A$ is the spherical (Bond) albedo and $r$ is the star--planet
separation. In this case the surface is uniformly bright and thus
there will be no observable phase function at infra-red
wavelengths. However, if the atmosphere is inefficient with respect to
heat redistribution, this will lead to a hot dayside for the planet
where the effective temperature is
\begin{equation}
  T_p = \left( \frac{L_\star (1-A)}{8 \pi \sigma r^2}
  \right)^\frac{1}{4}
\end{equation}
where there will be a resulting phase variation as the planet orbits the
star. The generalized form for the planetary effective temperature is thus
\begin{equation}
  T_p = \left( \frac{L_\star (1-A)}{(1 + \eta) 8 \pi \sigma r^2}
  \right)^\frac{1}{4}
  \label{plantemp}
\end{equation}
where $\eta$ is the atmospheric heat redistribution efficiency with a
value ranging between 0 and 1.

\begin{figure}
  \includegraphics[angle=270,width=8.2cm]{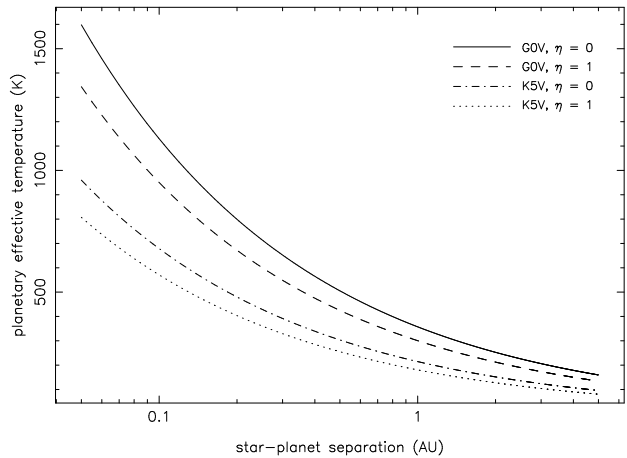}
  \caption{Dependence of planetary effective temperature on
    star--planet separation for inefficient ($\eta = 0$) and efficient
    ($\eta = 1$) heat redistributions. This is calculated for G0V and
    K5V host stars and shows that the scale of the temperature
    variations are dominated by the star--planet separation and not by
    the heat redistribution efficiency.}
  \label{sepdep}
\end{figure}

For a typical hot-Jupiter scenario, the star--planet separation is
assumed to be the same as the semi-major axis, $a$, since these are
usually circular orbits. However, the star--planet separation for
eccentric planets has the following form:
\begin{equation}
  r = \frac{a (1 - e^2)}{1 + e \cos f}
  \label{separation}
\end{equation}
where $e$ is the orbital eccentricity and $f$ is the true anomaly.
Thus, the eccentricity of a planetary orbit introduces a time
dependency to the effective temperature of the planet. In the absence
of atmospheric effects (see Section \ref{atmosphere}), the temperature
of the planet may be examined as indicated in Figure \ref{sepdep},
where the dependence on star--planet separation is shown for two
example spectral class targets. For a given spectral type, the
difference between 0\% and 100\% heat redistribution efficiency can
lead to a $\sim 20$\% adjustment in the temperature
calculation. However, $r$ is clearly dominant over $\eta$ at all
separations. This means that the planetary temperature, and thus the
flux ratio, is more dependent upon the orbital properties (which are
well-determined) than on the atmospheric dynamics (which are generally
unknown). This is elaborated upon in the following section.


\section{Atmospheric Properties}
\label{atmosphere}

Here we discuss the major atmospheric properties which directly
influence planetary thermal signatures.


\subsection{Spherical Albedo}

The spherical albedo of planetary atmospheres beyond the hot Jupiter
regime is less understood at IR wavelengths ($\sim 10$~$\mu$m) than it
is at optical wavelengths. However, the models of \citet{mar99} and
\citet{sud05} indicate that the albedo drops significantly in the
range between 1.0~$\mu$m and 1.2~$\mu$m in a manner which is
relatively independent of the star--planet separation. In the case of
Jupiter, the effective temperature of Jupiter beyond $\sim 5$~$\mu$m
is $\sim 125$~K and the spherical albedo is $\sim 60$\% of the
geometric albedo. In addition, constraints placed upon the spherical
albedos of hot Jupiters (e.g., \citet{knu09a}) show these to be
exceptionally low and in agreement with models which predict the
removal of reflective cloud layers in those extreme stellar flux
regimes. We thus generalize the albedo dependence on star--planet
separation adopted by \citet{kan10} by scaling the relation for
spherical albedos. However, here we are testing the conditions at
periastron where the albedo will be minimum. This is justified by the
low measured albedos mentioned above and the independent statistical
verification by \citet{cow11b} which favors low Bond albedos for small
star--planet separations.


\subsection{Heat Redistribution Efficiency}
\label{heat}

The small phase amplitude of HD~189733b detected by \citet{knu09a}
indicates that this particular planet has very high heat
redistribution efficiency caused by atmospheric advection which
produces high-speed zonal winds which carry heat to the night-side
of the planet. These atmospheric patterns are highly model dependent
\citep{bar07} and poorly understood for planets in the short-period
regime due to the small sample size and the complex interaction of
planetary structure, composition, tidal effects, and incident
flux. Longer period eccentric orbits such as HD~80606b will further
exhibit time-dependent behaviour depending upon the star--planet
separation and pseudo-synchronized spin rotation (see equation (42) of
\citet{hut81}). A statistical study of 24 known transiting planets
performed by \citet{cow11b} found that there is expected to be a wide
range in heat redistribution efficiencies. As described earlier, the
star--planet separation is dominant over heat redistribution
efficiency in determining the planetary effective
temperature. However, given the uncertainties in planetary models, we
consider the two extremes of $\eta = 0$ and $\eta = 1$ to determine
the upper and lower bounds on the flux ratio for a given planet.


\subsection{Radiative and Advective Time Constants}
\label{radadv}

The radiative time constant is a quantitative measure of the seasonal
lag caused by the thermal response of the atmosphere to incident flux
\citep{for08,sea05}. The radiative time constant,
$\tau_{\mathrm{rad}}$ is related to fundamental atmospheric properties
in the following way
\begin{equation}
  \tau_{\mathrm{rad}} \sim \frac{P}{g} \frac{c_P}{4 \sigma T_p^3}
  \label{tau_rad}
\end{equation}
where $P$ is the pressure, $g$ is the surface gravity, and $c_P$ is
the specific heat capacity.  If $\tau_{\mathrm{rad}} = 0$ then the
incident flux is immediately re-radiated from the day-side of the
planet \citep{cow11a}. For example, the measured value for HD~80606b by
\citet{lau09} is $\tau_{\mathrm{rad}} = 4.5 \pm 2$~hours.

The related quantity is the advective time constant,
$\tau_{\mathrm{adv}}$, which is a measure of the movement of a parcel
of gas around the planet. This is given by
\begin{equation}
  \tau_{\mathrm{adv}} \sim \frac{R_P}{U}
  \label{tau_adv}
\end{equation}
where $U$ is the characteristic wind speed. Thus,
$\tau_{\mathrm{adv}}$ approximates to zero when the wind speed becomes
large, a situation which results in a high heat redistribution
efficiency, as discussed in Section \ref{heat}. The predicted values
of $\tau_{\mathrm{rad}}$ and $\tau_{\mathrm{adv}}$ are highly
dependent upon the circulation models of the atmospheres
\citep{lan08,mon11,sho09} and also vary with the wavelength since this
effects the depth into which the atmosphere is probed
\citep{knu09a}. Here we concern ourselves with the peak flux ratio
which is expected to occur relatively close to the point of periastron
passage for eccentric orbits. Thus we assume that $\tau_{\mathrm{rad}}
\ll P$ and $\tau_{\mathrm{adv}} \gg P$ for the subsequent
calculations, keeping in mind that significant planetary spin, in
addition to atmospheric composition and cloud effects, may induce wind
patterns that cause divergence from this assumption. This produces an
upper limit on the flux from the planetary substellar point and thus
an upper limit on the predicted flux ratio with the star. The
interplay between the radiative/advective time constants and other
effects, such as drag mechanisms and numerical dissipation, have been
investigated by \citet{rau11,thr11}.


\section{Flux Ratio at Periastron Passage}

The measurable quantity from observations acquired at frequency $\nu$
is the flux ratio between the star and the planet, given by
\begin{equation}
  \frac{F_p}{F_\star} = \frac
       {(\exp{(h \nu / k T_\mathrm{eff}) - 1)} R_p^2}
       {(\exp{(h \nu / k T_p) - 1)} R_\star^2}
  \label{fluxratio}
\end{equation}
where $R_p$ is the radius of the planet.  The flux ratio as a function
of the orbital phase depends upon the thermal redistribution
efficiency of the atmosphere. If this is 100\% efficient then the flux
depends purely on the star--planet separation since there is no longer
a phase function such as that described by \citet{kan10}.

Observations of HD~189733b and HD~80606b indicate that these planets
have relatively high redistribution efficiency \citep{knu09a,lau09}.
These planets are in very different kinds of orbits and so this high
efficiency may be quite common amongst planets which experience either
constant or intermittent periods of high stellar flux. The lack of
atmospheric measurements, and thus models to explain observations, for
eccentric planets further motivates the need for this study. An
alternative hypothetical explanation is that hot Jupiters have high
zonal winds and very little cloud layer, whereas eccentric planets are
not irradiated in the same way and so retain some of their clouds,
even during periastron passage, thus reducing the redistribution
efficiency \citep{cow11a}. This could lead to phase variations in the
thermal signature such that planets whose periastron argument is
$\omega \sim 270\degr$ become the optimal targets. Discriminating
between these competing ideas requires further observations to
resolve. The calculations in the following sections thus represent
testable assumptions which can constrain these postulates.


\section{Habitability of Eccentric Planets}
\label{hz_sec}

The Habitable Zone (HZ) is defined as the range of circumstellar
distances from a star within which a planet could have liquid water on
its surface, given a dense enough atmosphere. The various criteria for
defining the HZ has been described in detail by \citet{kas93} and
further generalized as a function of spectral type by \citet{und03}
and \citet{jon10}. In estimating the boundaries of the HZ, we utlize
the equations of \citet{und03} which relate the radii of the HZ inner
and outer edges to the luminosity and effective temperature of the
host star. Using the boundary conditions of runaway greenhouse and
maximum greenhouse effects at the inner and outer edges of the HZ
respectively \citep{und03}, the stellar flux at these boundaries are
given by
\begin{eqnarray*}
  S_\mathrm{inner} = 4.190 \times 10^{-8} T_\mathrm{eff}^2 - 2.139
  \times 10^{-4} T_\mathrm{eff} + 1.268 \\
  S_\mathrm{outer} = 6.190 \times 10^{-9} T_\mathrm{eff}^2 - 1.319
  \times 10^{-5} T_\mathrm{eff} + 0.2341
\end{eqnarray*}
The inner and outer edgers of the HZ are then derived from the
following
\begin{eqnarray*}
  r_\mathrm{inner} = \sqrt{ L_\star / S_\mathrm{inner} } \\
  r_\mathrm{outer} = \sqrt{ L_\star / S_\mathrm{outer} }
\end{eqnarray*}
where the radii are in units of AU and the stellar luminosities are in
solar units.

\begin{figure}
  \includegraphics[width=8.5cm]{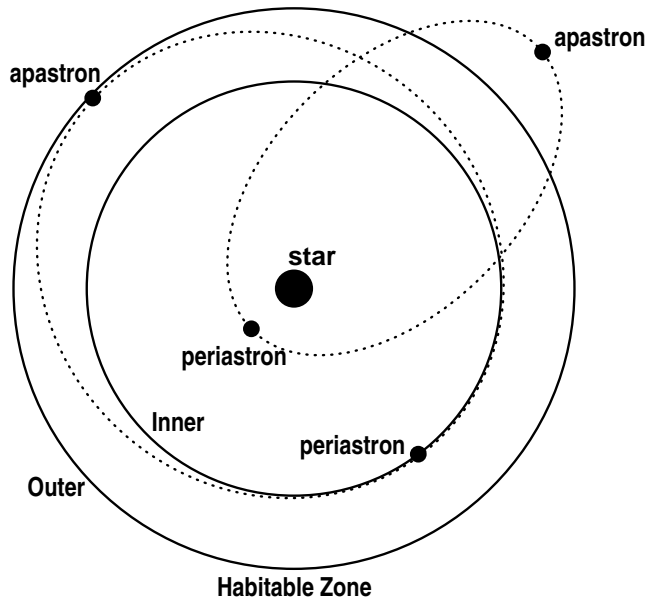}
  \caption{Eccentric orbits (dotted lines) relative to the inner and
    outer boundaries of the habitable zone (solid lines). The low
    eccentricity planet is able to maintain a presence inside the
    habitable zone, whilst the high eccentricity planet moves beyond
    both boundaries.}
  \label{hz_orbits}
\end{figure}

The effect of orbital eccentricity on the habitability of planets has
been investigated by \citet{ato04} and \citet{dre10}. Here we expand
upon this by using our effective temperature calculations to determine
the amount of time a planet spends in the HZ. There are two types of
habitable zone situations we consider: planets in eccentric orbits
that remain within the bounds of the habitable zone, and planets in
eccentric orbits that cross the boundaries of the habitable zone. In
the latter case, the atmospheric relaxation timescales (governed by
$\tau_{\mathrm{rad}}$ and $\tau_{\mathrm{adv}}$, described in Section
\ref{radadv}) may moderate temperatures even during dramatic changes
in star--planet separation. Note that long-period planets in circular
orbits are more likely to have a non-synchronous spin rotation and so
$\tau_{\mathrm{rad}}$ and $\tau_{\mathrm{adv}}$ will have a much
smaller correlation with orbital period. However, planets in eccentric
orbits will likely spend the majority of their orbits within the HZ if
apastron lies close to the outer edge, thus retaining the possability
of pseudo-synchronous spin rotation. Note that it is poorly understood
to what degree exoplanets in eccentric orbits will retain a
pseudo-synchronous spin rotation within the HZ due to the many
contributing factors. However, the evidence for these relative
influences have been observed for moons within our own Solar System
\citep{gla96}. Figure \ref{hz_orbits} shows two example eccentric
orbits overlaid on a hypothetical HZ which depicts the two cases
described above. In the following section, we apply these methods to
determine the percentage of the orbit spent in the HZ for the known
exoplanets.


\section{Application to Known Exoplanets}

Here we calculate predicted effective temperatures and flux ratios at
periastron passage for a sample of the known exoplanets for which
there are known orbital solutions. The orbital parameters of 390
planets, along with the host star properties, were extracted using the
Exoplanet Data Explorer\footnote{\tt http://exoplanets.org/}
\citep{wri11}. The data are current as of 17th January 2011. Planets
for which the host star values of $T_\mathrm{eff}$ and $\log g$ were
not available were excluded from the sample. The calculated flux
ratios are highly dependent on the assumed radii of the exoplanets
(see Equation \ref{fluxratio}. The anomlous radii of short-period
transiting planets has been investigated by \citet{lau11}. However,
\citet{for07} have shown that, for a given planetary mass and
composition, planetary radii should not vary substantially between
orbital radii of 0.1--2.0~AU. However, different compostions,
particularly with regards to core versus coreless models, can lead to
radii variations of $\sim 25$\%. Since we are considering eccentric
orbits of massive planets, we take the conservative approach of fixing
the radius for each of the planets in this sample at one Jupiter
radius, keeping the caveats mentioned above in mind.

There are a variety of current and future space-based observatories
that are capable of effectively monitoring known exoplanets for
thermal signatures. The IRAC instrument of the Spitzer Space Telescope
had passbands centered at 3.6, 4.5, 5.8, and 8.0
microns\footnote{http://ssc.spitzer.caltech.edu/irac/}. For the James
Webb Space Telescope (JWST), planned instruments include MIRI which
covers 5 to 27
microns\footnote{http://www.stsci.edu/jwst/instruments/miri/}, and
NIRCam which covers 0.6 to 5.0
microns\footnote{http://www.stsci.edu/jwst/instruments/nircam/}. We
calculate predicted flux ratios using the 3.6 and 4.5 micron passbands
of IRAC since the longer wavelengths are no longer available during
the warm phase of the mission.

\begin{deluxetable*}{lcccccccccc}
  \tablecolumns{11}
  \tablewidth{0pc}
  \tablecaption{\label{fluxtable} Effective temperatures and IR flux
    ratios for known exoplanets.}
  \tablehead{
    \multicolumn{5}{c}{} &
    \multicolumn{3}{c}{$\eta = 0$} &
    \multicolumn{3}{c}{$\eta = 1$} \\
    \colhead{Planet} &
    \colhead{$P$} &
    \colhead{$e$} &
    \colhead{$\omega$} &
    \colhead{$t_\mathrm{HZ}$} &
    \colhead{$T_p$} &
    \colhead{$F_p/F_\star$ ($10^{-3}$)} &
    \colhead{$F_p/F_\star$ ($10^{-3}$)} &
    \colhead{$T_p$} &
    \colhead{$F_p/F_\star$ ($10^{-3}$)} &
    \colhead{$F_p/F_\star$ ($10^{-3}$)} \\
    \colhead{} & \colhead{(days)} & \colhead{} &
    \colhead{($\degr$)} & \colhead{(\%)} &
    \colhead{(K)} & \colhead{(3.6~$\mu$m)} & \colhead{(4.5~$\mu$m)} &
    \colhead{(K)} & \colhead{(3.6~$\mu$m)} & \colhead{(4.5~$\mu$m)}
  }
  \startdata
HD 80606 b      &  111.4 & 0.934 & 300.6 &  39.7 & 1838.2 & 1.5762187 & 1.9343872 & 1545.7 &  1.0012931 & 1.3134501 \\
HD 20782 b      &  585.9 & 0.925 & 147.0 &  22.8 & 1080.4 & 0.2262089 & 0.3613180 &  908.5 &  0.1109253 & 0.2016740 \\
HD 4113 b       &  526.6 & 0.903 & 317.7 &  23.4 &  946.0 & 0.1481725 & 0.2604596 &  795.5 &  0.0660831 & 0.1351611 \\
HD 156846 b     &  359.5 & 0.847 &  52.2 &  60.5 & 1098.9 & 0.0946147 & 0.1502021 &  924.0 &  0.0469136 & 0.0845516 \\
HD 43197 b      &  327.8 & 0.830 & 251.0 &  77.9 &  851.5 & 0.0808349 & 0.1547891 &  716.0 &  0.0330801 & 0.0751514 \\
HD 28254 b      & 1116.0 & 0.810 & 301.0 &  18.6 &  619.3 & 0.0077581 & 0.0209590 &  520.7 &  0.0022853 & 0.0078623 \\
HD 45350 b      &  963.6 & 0.778 & 343.4 &  15.0 &  529.9 & 0.0042552 & 0.0142493 &  445.6 &  0.0010208 & 0.0045419 \\
HD 30562 b      & 1157.0 & 0.760 &  81.0 &  18.4 &  543.9 & 0.0031772 & 0.0102831 &  457.4 &  0.0007908 & 0.0033750 \\
HD 20868 b      &  380.9 & 0.750 & 356.2 &  36.6 &  601.7 & 0.0139532 & 0.0385124 &  506.0 &  0.0039670 & 0.0140474 \\
HD 37605 b      &   54.2 & 0.737 & 211.6 &   0.0 & 1160.8 & 0.4103989 & 0.6223644 &  976.1 &  0.2106124 & 0.3596571 \\
HD 222582 b     &  572.4 & 0.725 & 319.0 &  30.6 &  567.1 & 0.0077235 & 0.0234816 &  476.9 &  0.0020345 & 0.0080621 \\
HD 8673 b       & 1634.0 & 0.723 & 323.4 &  16.6 &  487.9 & 0.0011613 & 0.0044675 &  410.3 &  0.0002465 & 0.0012917 \\
HD 2039 b       & 1120.0 & 0.715 & 344.1 &  15.0 &  456.8 & 0.0012679 & 0.0054234 &  384.1 &  0.0002422 & 0.0014417 \\
HD 96167 b      &  498.9 & 0.710 & 285.0 &  60.9 &  745.3 & 0.0126504 & 0.0276510 &  626.7 &  0.0045728 & 0.0121864 \\
HD 86264 b      & 1475.0 & 0.700 & 306.0 &  19.6 &  487.1 & 0.0010770 & 0.0041537 &  409.6 &  0.0002280 & 0.0011985 \\
HAT-P-13 c      &  428.5 & 0.691 & 176.7 &  68.9 &  667.8 & 0.0111062 & 0.0273530 &  561.5 &  0.0035730 & 0.0110002 \\
HD 159868 b     &  986.0 & 0.690 &  97.0 &  30.0 &  523.3 & 0.0019888 & 0.0067876 &  440.0 &  0.0004686 & 0.0021329 \\
HD 17156 b      &   21.2 & 0.682 & 121.9 &   0.0 & 1767.7 & 0.6833214 & 0.8563483 & 1486.4 &  0.4281503 & 0.5754836 \\
16 Cyg B b      &  798.5 & 0.681 &  85.8 &  20.4 &  463.6 & 0.0016843 & 0.0069971 &  389.8 &  0.0003295 & 0.0018960 \\
HD 89744 b      &  256.8 & 0.673 & 195.1 &   0.0 &  907.5 & 0.0321474 & 0.0589155 &  763.1 &  0.0138748 & 0.0298084 \\
HD 39091 b      & 2151.0 & 0.641 & 330.2 &  11.9 &  330.3 & 0.0000443 & 0.0003698 &  277.8 &  0.0000045 & 0.0000592 \\
HD 131664 b     & 1951.0 & 0.638 & 149.7 &  11.1 &  322.1 & 0.0000384 & 0.0003408 &  270.9 &  0.0000037 & 0.0000521 \\
HD 74156 b      &   52.0 & 0.630 & 174.0 &   0.0 & 1235.3 & 0.2161417 & 0.3191066 & 1038.7 &  0.1150297 & 0.1895808 \\
HD 44219 b      &  472.3 & 0.610 & 147.4 &  73.9 &  552.5 & 0.0045355 & 0.0143141 &  464.6 &  0.0011534 & 0.0047790 \\
HD 154672 b     &  163.9 & 0.610 & 265.0 &   0.0 &  766.4 & 0.0358256 & 0.0760486 &  644.4 &  0.0133102 & 0.0342433 \\
HD 16175 b      &  990.0 & 0.600 & 222.0 &  23.8 &  419.6 & 0.0004756 & 0.0023800 &  352.8 &  0.0000784 & 0.0005626 \\
HD 3651 b       &   62.2 & 0.596 & 245.5 &   0.0 &  833.3 & 0.1188941 & 0.2310287 &  700.7 &  0.0477421 & 0.1105272 \\
HD 171028 b     &  550.0 & 0.590 & 304.0 &  54.7 &  623.3 & 0.0045460 & 0.0121809 &  524.1 &  0.0013496 & 0.0045978 \\
HIP 2247 b      &  655.6 & 0.540 & 112.2 &  25.4 &  353.8 & 0.0001572 & 0.0010949 &  297.5 &  0.0000185 & 0.0001980 \\
HD 190228 b     & 1136.1 & 0.531 & 101.2 &  40.0 &  420.4 & 0.0001656 & 0.0008165 &  353.5 &  0.0000274 & 0.0001936 \\
CoRoT-10 b      &   13.2 & 0.530 & 218.9 &   0.0 & 1123.2 & 0.6402574 & 0.9856290 &  944.5 &  0.3219422 & 0.5607959 \\
HD 142022 b     & 1928.0 & 0.530 & 170.0 &   8.3 &  273.5 & 0.0000047 & 0.0000650 &  230.0 &  0.0000003 & 0.0000071 \\
HD 87883 b      & 2754.0 & 0.530 & 291.0 &   0.0 &  193.4 & 0.0000000 & 0.0000010 &  162.6 &  0.0000000 & 0.0000000 \\
HD 108147 b     &   10.9 & 0.530 & 308.0 &   0.0 & 1828.5 & 0.7475883 & 0.9266767 & 1537.6 &  0.4740451 & 0.6283575 \\
HD 168443 b     &   58.1 & 0.529 & 172.9 &   0.0 &  959.0 & 0.1119503 & 0.1943817 &  806.4 &  0.0504508 & 0.1016838 \\
HD 81040 b      & 1001.7 & 0.526 &  81.3 &  20.3 &  325.6 & 0.0000604 & 0.0005207 &  273.8 &  0.0000059 & 0.0000813 \\
HIP 5158 b      &  345.7 & 0.520 & 252.0 &  49.8 &  425.7 & 0.0012029 & 0.0057508 &  358.0 &  0.0002036 & 0.0013882 \\
HD 148156 b     & 1027.0 & 0.520 &  35.0 &  24.9 &  350.4 & 0.0001130 & 0.0008256 &  294.6 &  0.0000131 & 0.0001469 \\
HD 217107 c     & 4270.0 & 0.517 & 198.6 &   0.0 &  197.0 & 0.0000000 & 0.0000008 &  165.6 &  0.0000000 & 0.0000000 \\
HAT-P-2 b       &    5.6 & 0.517 & 185.2 &   0.0 & 2498.6 & 1.2828619 & 1.4573069 & 2101.0 &  0.8890871 & 1.0564045 \\
HD 1237 b       &  133.7 & 0.511 & 290.7 &  42.5 &  585.8 & 0.0179584 & 0.0520765 &  492.6 &  0.0049349 & 0.0184897 \\
HD 142415 b     &  386.3 & 0.500 & 255.0 &  72.3 &  487.3 & 0.0023210 & 0.0088971 &  409.8 &  0.0004917 & 0.0025685 \\
HD 215497 c     &  567.9 & 0.490 &  45.0 &  35.6 &  376.9 & 0.0002662 & 0.0016269 &  317.0 &  0.0000358 & 0.0003268 \\
HD 106252 b     & 1531.0 & 0.482 & 292.8 &  17.1 &  302.4 & 0.0000157 & 0.0001642 &  254.3 &  0.0000013 & 0.0000222 \\
HD 33636 b      & 2127.7 & 0.481 & 339.5 &   4.1 &  262.2 & 0.0000024 & 0.0000372 &  220.5 &  0.0000001 & 0.0000037 \\
HD 181433 d     & 2172.0 & 0.480 & 330.0 &   0.0 &  221.7 & 0.0000002 & 0.0000057 &  186.4 &  0.0000000 & 0.0000004 \\
HD 196885 b     & 1333.0 & 0.480 &  78.0 &  35.9 &  366.0 & 0.0000877 & 0.0005804 &  307.8 &  0.0000111 & 0.0001112 \\
HD 33283 b      &   18.2 & 0.480 & 155.8 &   0.0 & 1503.7 & 0.3606475 & 0.4817197 & 1264.5 &  0.2118062 & 0.3083335 \\
HD 210277 b     &  442.2 & 0.476 & 119.1 &  44.6 &  398.9 & 0.0005740 & 0.0031455 &  335.5 &  0.0000862 & 0.0006903 \\
HD 154857 b     &  409.0 & 0.470 &  59.0 &  32.1 &  575.2 & 0.0021955 & 0.0065333 &  483.7 &  0.0005893 & 0.0022767 \\
  \enddata
  \tablecomments{The last six columns are for the cases of $\eta = 0$
    and $\eta = 1$ respectively.}
\end{deluxetable*}

\begin{figure*}
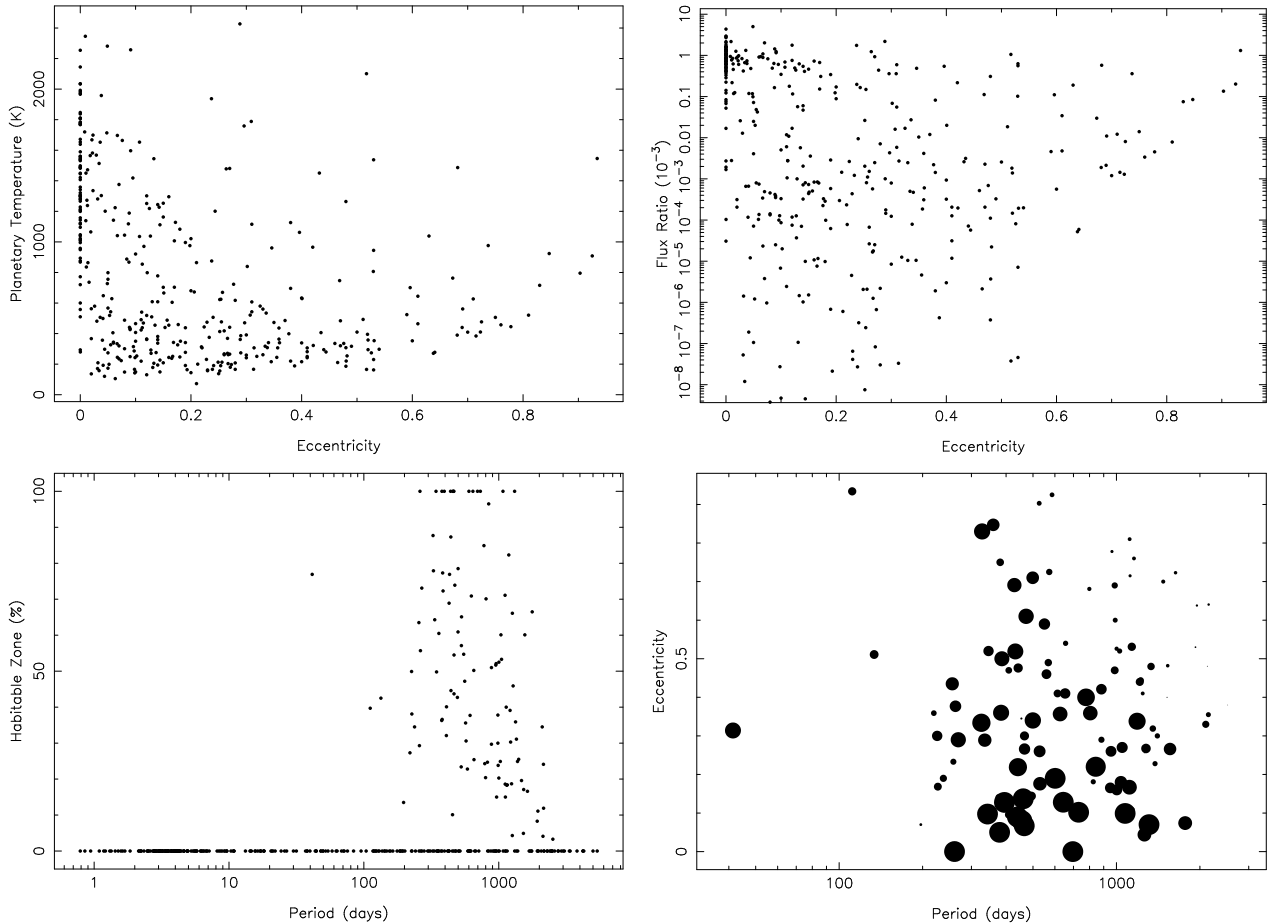

  \begin{center}
    \begin{tabular}{cc}
      \includegraphics[angle=270,width=8.2cm]{f03a.ps} &
      \includegraphics[angle=270,width=8.2cm]{f03b.ps} \\
      \includegraphics[angle=270,width=8.2cm]{f03c.ps} &
      \includegraphics[angle=270,width=8.2cm]{f03d.ps} \\
    \end{tabular}
  \end{center}
  \caption{Predicted planetary effective temperature at periastron
    (top-left) and predicted flux ratio at 4.5~$\mu$m (top-right) as a
    function of eccentricity. The predicted flux ratios assume a
    well-mixed ($\eta = 1$) atmospheric model and an albedo of $A =
    0.0$. Also shown are the the percentage of the orbit spent within
    the HZ with respect to orbital period (bottom-left) and the
    eccentricity as a function of period (bottom-right). The last
    panel only shows planets which enter the HZ where the size of the
    points linearly increases with the percentage of time spent within
    the HZ.}
  \label{fluxecc}
\end{figure*}

Table \ref{fluxtable} shows the results of these calculations for the
50 most eccentric planets in the sample. Included in the table are the
orbital period, $P$, eccentricity, $e$, periastron argument, $\omega$,
the percentage of a full orbit spent in the HZ, $t_\mathrm{HZ}$, and
the effective temperatures and IR flux ratios assuming both low ($\eta
= 0$) and high ($\eta = 1$) heat redistribution efficiencies. We also
assume an albedo of $A = 0.0$, meaning that the planet absorbs 100\%
of the incident flux (see \citet{mad09} for a more thorough
statistical analysis of these values). This produces a higher
effective temperature for the planet but is a reasonable assumption
since models and measurements have shown that planets lose their
reflective cloud layers at small star--planet separations
\citep{kan10,sud05}. For comparison, an albedo of $A = 0.3$ leads to a
$\sim 10$\% reduction in the planetary effective temperatures. Note
that flux ratios for those planets at particularly long periods are
approximately zero at both passbands. However, recall from Figure
\ref{sepdep} that the flux ratios are mostly dependent upon the
star--planet separation at periastron rather than $\eta$. Thus, as one
increases the orbital period and decreases the eccentricity, the
change in $\eta$ becomes less important (also demonstrated later in
the top-right panel of Figure \ref{fluxecc}). The flux ratios
calculated for HD~80606b are comparable to the $0.0010 \pm 0.0002$
values measured by \citet{lau09} at 8~$\mu$m and indicate that there
is moderate heat redistribution efficiency of the atmosphere in this
case. Some of the more interesting examples are discussed in detail in
Section \ref{case}.

The top two panels of Figure \ref{fluxecc} plot the planetary
effective temperatures and flux values for all 390 planets included in
the sample, assuming $\eta = 1$ and $\lambda = 4.5$~$\mu$m. The
planetary temperature plot on the left shows a reasonably even
distribution of temperatures for circular orbits, as one may
expect. However, notice there is a downward trend for eccentricities
$\lesssim 0.5$ since these planets tend to lie at a large semi-major
axis. A second trend occurs in the opposite direction (eccentricities
$\gtrsim 0.5$); the temperature increases as the eccentricity
increases since this decreases the star--planet separation at
periastron. The temperature plot maps to the flux ratio plot shown on
the right, where the hot Jupiters can be seen clustered in the
top-left corner. The upward trend in temperatures towards higher
eccentricities results in an equivalent upward trend in flux
ratios. There is no decrease in flux ratio towards higher
eccentricities. Thus it is clear that the high eccentricity planets
present viable opportunities to detect their thermal signature during
periastron passage.

The percentage of the total orbital period spent within the HZ is
calculated by first estimating the boundaries of the HZ (see Section
\ref{hz_sec}) and then determining the star--planet separation at
equal increments in time during a Keplerian orbit. It is not
surprising that most of the eccentric planets spend less than half of
their time within the HZ due to the large range of star--planet
separations which occur during the orbit. The bottom two panels of
Figure \ref{fluxecc} show how the planets which spend some part of
their orbit within the HZ are distributed according to period and
eccentricity. The left plot shows the percentage of the orbit spent in
the HZ, $t_\mathrm{HZ}$, as a function of period. Interestingly, the
planets which spend more than 20\% of the orbit within the HZ are
event distributed between 20--80\% and orbital periods of
200--2000~days. To clarify the relative distributions of the planets
which spend a non-zero amount of time within the HZ, we show the
eccentricity of the planets as a function of period in the right
plot. The relative size of the points increases as a function of the
percentage time spent within the HZ. The planets which spend a portion
of their time in the HZ are fairly evenly distributed in eccentricity,
although the more circular orbits preferentially spend a greater
percentage of their time there. This indicates eccentricity can
sometimes be a useful discriminator in selecting targets for potential
life-bearing planets, although there are some relatively large points
shown for $e > 0.5$. In both of the bottom plots, the outlier located
at 41.4 days is the Saturn-mass planet orbiting the M4 dwarf
HIP~57050, discovered by \citet{hag10}. Despite the relatively short
period and eccentricity of 0.31, the smaller luminosity of the host
star allows this planet to spend most of the orbit within the HZ.

\begin{figure*}
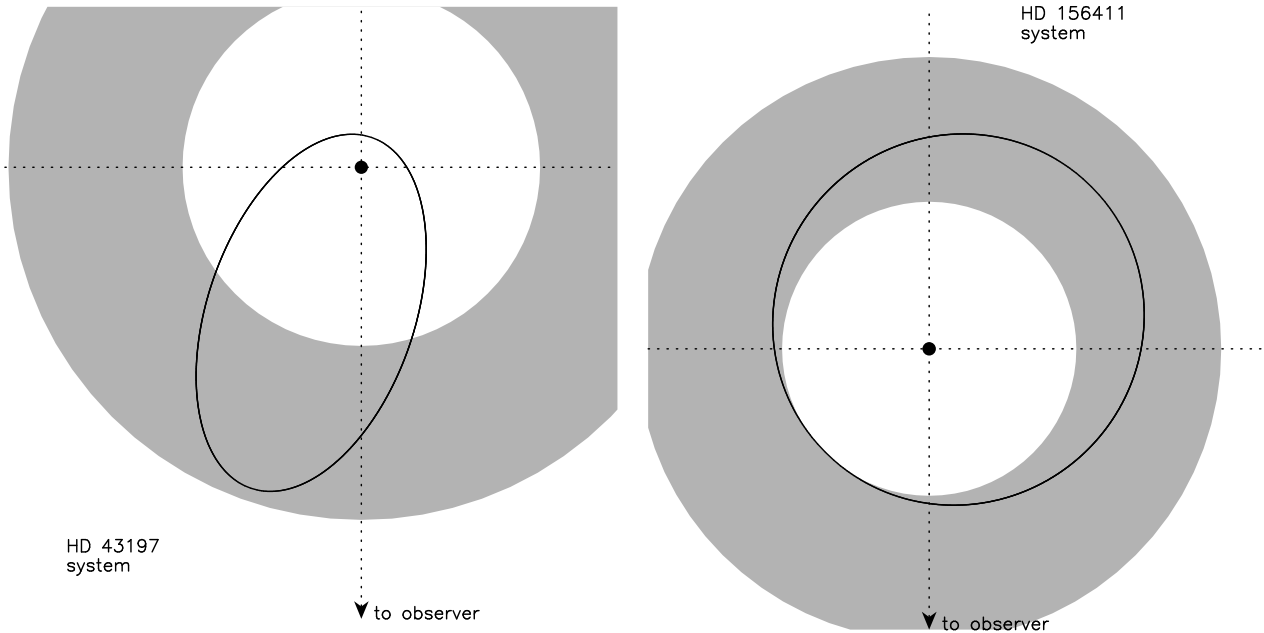

  \begin{center}
    \begin{tabular}{cc}
      \includegraphics[angle=270,width=8.2cm]{f04a.ps} &
      \includegraphics[angle=270,width=8.2cm]{f04b.ps}
    \end{tabular}
  \end{center}
  \caption{Habitable zones (shaded region) and orbits of the planets
    orbiting HD~43197 (left) and HD~156411 (right).}
  \label{hzplots}
\end{figure*}

In Figure \ref{hzplots} we show two particularly interesting cases. On
the left is shown the highly eccentric planet orbiting HD~43197. This
planet spends $\sim 78$\% of the total orbital period residing within
the HZ, although even considering $\eta = 1$, the temperature rises to
716~K during periastron passage. On the right is shown the rather more
benign orbit of the planet orbiting HD~156411. This planet is not
listed in Table \ref{fluxtable} (due to its relatively low
eccentricity) but it spends 96.5\% of the orbit within the confines of
the HZ and reaches a peak temperature of 307~K during periastron
passage assuming $\eta = 1$. With a flux ratio of $7.8 \times 10^{-8}$
at 4.5~$\mu$m, one cannot reasonably expect to detect such a planet
with current instrumentation, but it does present an interesting case
for habitability studies of eccentric orbits. These two planets
perfectly represent the two cases described by Figure \ref{hz_orbits}.

\section{Impact of Star Spots on Detection}

Since most of the known exoplanets we are concerned with here orbit F,
G, or K-type main sequence stars, we need to investigate the effect,
if any, that star spots will have on the detection of these planets
and their possible phase signatures in the IR. \citet{ber05} showed
that, on average, the difference between the stellar photosphere and a
stellar spot is larger for hotter stars with values near 2000~K for
late F and early G stars and drops to 200~K for late M stars.  IR
observations detect stars at cooler temperatures than optical
observations, and therefore star spots will be observed at a lower
contrast than in the optical. This means that, even though late-type
stars typically have a greater number of star spots, these should be
less of a flux contrast issue for the late-type stars that are better
suited for IR observations.

Of concern is the timescale of star spot modulations compared with the
timescale over which significant flux variation is expected to occur
during periastron passage. As shown by \cite{kan10}, the maximum flux
amplitude for eccentric planets is a small fraction of the total orbit
and in many cases allows a 24--48 hour window through which the
maximum changes can be observed. In contrast, the rotation periods of
typical exoplanet hosting stars is substantially longer. Rotation
periods have been measured for many of these, such as the work of
\citet{hen97} and \citet{sim10}. These rotation periods are mostly in
the range of 20--40 days which is common of 1--5~Gyr old main sequence
stars.

In the case of the sun, star spots typically cover between 10$^{-3}$
(during a solar maximum) and 10$^{-5}$ of the surface. However, up to
a 22\% of a hemisphere was seen to be covered in a Doppler image of XX
Tri, which is a K0 giant and therefore not typical of exoplanet host
stars \citep{str09}. Star spots are likely to evolve over timescales
of a few stellar rotation cycles or even within one cycle. This
supports the results of \citet{hus02} who found that spots on single
main sequence stars, at most, live for weeks.

The frequency of star spot occurrance increases for low-mass stars
which, at the current epoch, comprise the minority of exoplanet host
stars. As described above, most stellar rotation periods are much
longer than the periastron passage timescales. Therefore it is
usuaully not necessary to worry about star spots inhibiting the
detection of the phase signatures of the planets. Stellar spot
signatures change over time, so in the very few cases where the
timescales may be similar, it should be possible to disentangle the
planet phase signature from the star spot signature.  Consistent with
our findings, an IR study conducted by \citet{des11a} found that the
variability due to spots is less than the predicted transit depths and
with a longer period. Conversely, \citet{des11b} find that they do
have to take into account spots on HD~189733 and conclude that an
estimation of the planet-to-star radius ratio should be associated
with a corresponding stellar activity level. Thus, each experiment
should address this stellar activity level issue on a case-by-case
basis.


\section{Case Studies}
\label{case}

In this section we consider several interesting case studies from the
results in the previous sections.


\subsection{The HD~156846 System}

The planet orbiting HD~156846 has recently been studied by
\citet{kan11b}, providing refined orbital parameters and the exclusion
of additional giant planets in the system. Thus, even though the
orbital period is large ($P = 359$~days), observations during
periastron passage could be timed with great precision in order to
detect a thermal signature. However, the periastron argument of
$\omega = 51\degr$ means that the periastron passage occurs close to
the observer--star line of sight on the near side of the star. Thus
the night-side of the planet faces towards us during periastron
passage. This presents an opportunity to test the heat redistribution
efficiency models for this planet since a low efficiency will result
in a distinct phase function which prevents detection, but a high
efficiency will still allow for a detectable signature as the heat is
transferred to the night-side of the planet.


\subsection{The HD~37605 and HD~33283 Systems}

\begin{figure*}
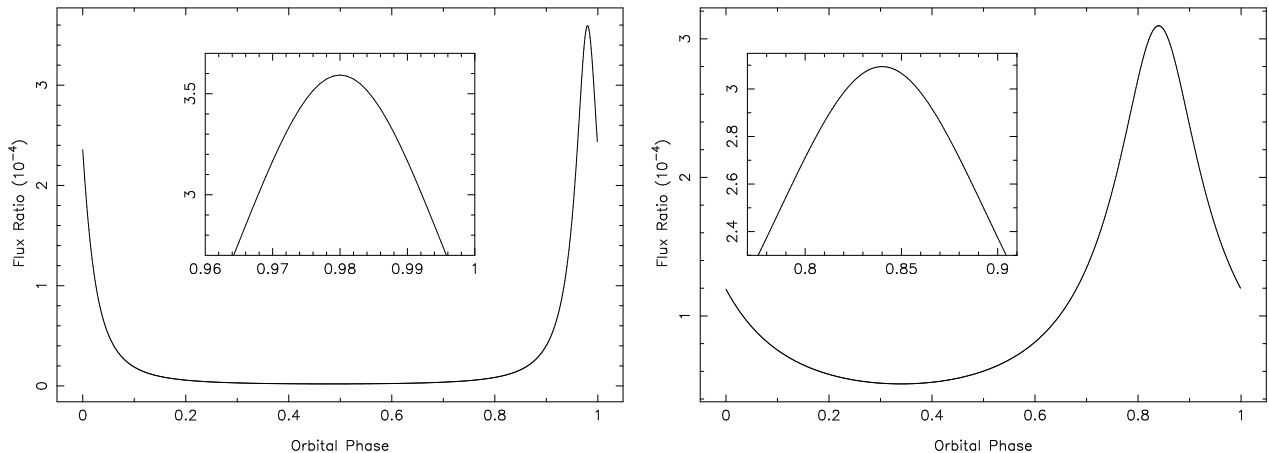

  \begin{center}
    \begin{tabular}{cc}
      \includegraphics[angle=270,width=8.2cm]{f05a.ps} &
      \includegraphics[angle=270,width=8.2cm]{f05b.ps}
    \end{tabular}
  \end{center}
  \caption{Phase curves for HD~37605b (left) and HD~33283b (right) at
    4.5~$\mu$m, assuming 100\% heat redistribution efficiency ($\eta =
    1$) of the atmosphere. The sub-panels zoom in on the section of
    the phase curve which would be optimal for monitoring.}
  \label{phasecurves}
\end{figure*}

Here we discuss the two interesting cases of HD~37605b and
HD~33283b. The flux ratio of HD~37605b is second only to HD~80606b
amongst the eccentric planets shown in Table \ref{fluxtable}, aided by
the short orbital period compared with its large eccentricity. This
planet was also suggested as an excellent candidate for phase
monitoring at optical wavelengths by \citet{kan10}. The periastron
argument of $\omega = 211\degr$ means that a significant fraction of
the day-side will be angled towards us during periastron passage.
HD~33283b has a very similar predicted flux ratio, the lower
eccentricity being offset by a shorter orbital period.  The host stars
in both cases are bright enough such that a detection of phase
variations during periastron passage will yield significant
information regarding the atmospheric properties of the respective
planets. A lack of detectable phase variation will in turn constrain
atmospheric models for these planets regarding the mechanisms for
which heat distribution is occurring.

Provided the orbital parameters for these stars are sufficiently
refined, the total time needed to achieve adequate coverage of the
phase curves to maximize a detection is relatively low. Shown in
Figure \ref{phasecurves} are the calculated flux ratios at 4.5~$\mu$m
as a function of orbital phase for each of the planets. Zero orbital
phase is defined to occur at superior conjunction and the location of
the peak flux ratio depends upon the periastron argument of the
orbit. This model assumes that the atmospheres are very efficient at
redistributing the heat $\eta = 1$ which lowers the effective flux
ratios. Thus, these models represent a minimum expected signature from
each of the planets. Each of the figures contains a sub-panel which
zooms in on the pericenter passage segment of the orbit. These
segments can be adequately covered with 25 hours of observations,
therefore only 50 hours in total would be needed to monitor both
targets. The planets have the same time-coverage requirements because,
although HD~37605b has a longer period than HD~33283b, it also has a
larger eccentricity which increases the planetary velocity at
pericenter. Such targets represent rare opportunities to probe the
atmospheric properties of planets beyond the regime of hot Jupiters
and further develop theoretical models of these planets.


\section{Conclusions}

We have presented an analysis of the known exoplanets to determine
predicted temperatures and flux ratios. This exercise is clearly not
meant to serve as an exhaustive modeling of the atmospheres for each
of these planets and indeed combining the models of \citet{for10} and
\citet{lew10} with IR data will present new and exciting opportunities
for studying the atmospheres of eccentric planets. Rather, this is
meant to serve as a guide towards which of the particularly eccentric
exoplanets may serve as interesting targets for follow-up
observations.

For planets which lie within the HZ of their host stars, eccentric
planets may present additional opportunities for studying planetary
atmospheres in these zones. HD~43197b and HD~156411b are particularly
interesting cases since they each spend a majority of the total
orbital period within their respective habitable zones but vary
dramatically in the temperature differences experienced throughout
their entire orbits. Thus these cases may be used to investigate the
effects of eccentric orbits upon habitability. Based on the fact that
most stellar rotation periods are are much longer than the periastron
passage timescales, and that star spots typically have a signature
that is much less than that of a habitable zone planet at periastron
passage, we conclude that spots are not a significant issue to this
particular study.  Due to the changing nature of star spot activity on
a given star, long-term monitoring of the star could disentangle the
planet phase signature.

We have presented several case studies from our analysis which serve
as potentially interesting targets for detection during periastron
passage. HD~156846b is a good IR target for studies of the atmospheric
properties of planets in extreme (highly eccentric) orbits. HD~37605b
and HD~33283b are also good targets since they have bright host stars,
relatively high predicted flux ratios, and brief periods of large
changes in the flux ratio. Warm Spitzer is capable of monitoring some
of these target during the remaining lifetime of the mission. When
this mission ceases operations, JWST will have instrumentation that is
capable of continuing this study to even higher precision.


\section*{Acknowledgements}

The authors would like to thank Ian Crossfield, Gregory Laughlin,
Nicolas Cowan, and Lisa Prato for several useful discussions. We would
also like to thank the anonymous referee, whose comments greatly
improved the quality of the paper. This research has made use of the
Exoplanet Orbit Database and the Exoplanet Data Explorer at
exoplanets.org.


\end{document}